\documentclass[aps,superscriptaddress,showpacs]{revtex4}
\usepackage{amssymb}

\usepackage{amsmath}
\usepackage{graphicx,epsfig}

\begin{document}

\title{Random Walk Access Times on Partially-Disordered Complex Networks:\\ an
Effective Medium Theory}
\author{Paul E. Parris}
\affiliation{Consortium of the Americas for Interdisciplinary Science and Department of\\
Physics and Astronomy, University of New Mexico, Albuquerque, NM 87131, USA}
\affiliation{Department of Physics, University of Missouri-Rolla, Rolla, MO 65409, USA}
\author{Juli\'an Candia}
\affiliation{Consortium of the Americas for Interdisciplinary Science and Department of\\
Physics and Astronomy, University of New Mexico, Albuquerque, NM 87131, USA}
\affiliation{Center for Complex Network Research and Department of Physics,\\
University of Notre Dame, Notre Dame, IN 46556, USA}
\author{V.M. Kenkre}
\affiliation{Consortium of the Americas for Interdisciplinary Science and Department of\\
Physics and Astronomy, University of New Mexico, Albuquerque, NM 87131, USA}

\begin{abstract}
An analytic effective medium theory is constructed to study the mean access
times for random walks on hybrid disordered structures formed by embedding
complex networks into regular lattices, considering transition rates $F$
that are different for steps across lattice bonds from the rates $f$ across
network shortcuts. The theory is developed for structures with arbitrary
shortcut distributions and applied to a class of partially-disordered
traversal enhanced networks in which shortcuts of fixed length are
distributed randomly with finite probability. Numerical simulations are
found to be in excellent agreement with predictions of the
effective medium theory on all aspects addressed by the latter. Access times
for random walks on these partially disordered structures are compared to
those on small-world networks, which on average appear to provide the most
effective means of decreasing access times uniformly across the network.
\end{abstract}


\maketitle

\section{Introduction}

There has been considerable recent interest in the statistical, topological,
and dynamical properties of complex networks, important examples of which
include the well-known scale-free and small-world networks (SFNs and SWNs).
Justification for this interest stems from the increasing use of complex
network theory to the understanding of a wide variety of phenomena in the
social, biological, and physical sciences, a small sample of which include
collaboration and acquaintance networks, protein interactions in the cell,
power grids, entangled polymers, and information and communication networks %
\cite{wat98,nw,wat99,new03,alb02,dor03,dor02a,kuperman,rm,candia}. In
addition to the ongoing interest in static topological properties of such
complex, generally disordered structures, there has been a parallel interest
in various \emph{dynamical} processes defined on them, and on how such
processes differ from those that take place on ordered networks and regular
lattices (RLs) \cite%
{sca01,pan01,lah01,alm03,blu,noh04,yan05,cos07,bo,derek,par05,can07,kor07}.

In particular, a recent contribution to the study of random walks on
Newman-Watts SWNs, which are structures formed by superimposing a classical
random graph of randomly distributed links onto the sites of an ordered
regular lattice \cite{wat98,nw,wat99,new03}, is a variation introduced by
Parris and Kenkre \cite{par05} in which the random walk proceeds in
continuous rather than discrete time, and the jump rate $f$ for steps across
the random shortcuts is different from the rate $F\ $for steps across
regular lattice bonds. Such a modification has two primary advantages.
First, it allows the strength of the regular and random graph components of
the network to be independently adjusted (or even turned off) at fixed
shortcut density by simply varying the ratio $f/F$. Second, it allows for
the rational evaluation of schemes for upgrading or optimizing access times
in existing information or communication networks, e.g., by randomly
incorporating a relatively small number of newer, faster connections into an
existing network already possessing a large number of older, perhaps slower,
ones.

The analysis of Ref.~\cite{par05} focused on the network \textit{traversal
time }$\tau =\tau _{N/2}$, defined as the mean first-passage time required
by a random walker initially at site $n$ to reach the site at $m=n+N/2,$
farthest from where it started on the regular 1D lattice backbone (formed
into a ring of $N$ nodes), averaged over initial sites and network
realizations. That study revealed an interesting collapse of traversal time
data onto a universal curve for fast shortcuts $f/F\gg 1$ and low shortcut
densities $n_{sw}=N_{sw}/N=2\bar{k},$ defined as the ratio of the number $%
N_{sw}$ of small-world shortcuts to the total number $N$ of sites (or nodes)
in the system. This ratio $n_{sw}$ is equal to twice the average \emph{shortcut vertex degree} $\bar{k}.$

The present paper explores a question that has naturally arisen from
an extension \cite{can07} of the work of Ref. \cite{par05} on SWNs to the
study of random walks on hybrid structures
formed by superimposing \emph{scale-free} shortcut networks onto regular
lattices, carried out recently considering
more general \emph{access times} $\tau _ {m}$
for a walker to reach a location $m$ nodes away from where it started. One
conclusion of Ref.~\cite{can07} was that, at fixed shortcut degree $\bar{k}$
and shortcut speed $f,$ SFN-RL hybrids are always less efficient at reducing
the mean traversal time $\tau ,$ and result in more dispersion in the
distribution of access times across the network, than their SWN
counterparts. This conclusion, which simply reflects the more inhomogeneous
nature of local environments in SFNs relative to SWNs, raises the question as
to whether there exist other random structures whose embedding would
generally enhance network traversal and access times more efficiently than
SWNs.

To explore this question, we consider a
new class of partially-ordered random structures that are generally more
ordered than the SWNs studied in Refs.~\cite{par05} and \cite{can07}, and
that we generically refer to as \emph{traversal enhanced networks} (TENs).
Motivation for the development of such structures arises from the intuition
that a walker's ability to traverse the system is not efficiently enhanced
by shortcuts between nodes that are already close together. Indeed, in a
given SWN system, if some of the shorter random links could be redirected
into nodes farther away, the mean traversal and access times for the entire
structure would presumably decrease without changing the total number of
shortcuts.

We consider here, therefore, random walks on a 1D ring of $N$ nodes
connected by nearest-neighbor hopping rate $F,$ and associate with each
possible link between two nodes separated by a backbone distance $\left|
n\right| \leq N/2$ an \textit{a priori} shortcut connection probability $%
q_{n}=q_{-n},$ with $1\geq q_{n}\geq 0,$ such that $q_{0}=q_{1}=0,$ and $%
\bar{k}=\sum_{n}2q_{n}.$ Two nodes separated by a particular backbone span $%
\left| n\right| $ are connected by a hopping rate $f$ with probability $q_{n}
$ and are left unconnected with probability $1-q_{n}$. The SWN hybrids
previously studied in Ref.~\cite{par05} are seen to be limiting cases of
such a network with \emph{equal} connection probabilities $q_{n}=q\sim \bar{k%
}/2N$. To optimize network traversal times, or times for access to an
arbitrary site, on the other hand, TENs allow for the possibility of
increasing connection probabilities $q_{n}$ for large $\left| n\right| $
relative to those for small $\left| n\right| ,$ keeping fixed the total
average shortcut degree $\bar{k}$. The general theory that we develop in
this paper allows for the study of partially ordered networks with an
arbitrary shortcut connection probability distribution $q_{n}$. For purposes
of comparison with simulation, however, the main results of the present
paper are applied to a specific subclass of such structures for which $%
q_{n}=q\delta _{\left| n\right| ,n_{0}},$ i.e., for which shortcuts of a
single span $n_{0}$ are randomly added. As might be expected, adding
shortcuts of length $n_{0}$ dramatically decreases access times for pairs of
sites separated by that distance. Perhaps less intuitively, however, we also
find that the \emph{average} access time to any point on the network
fluctuates rapidly with $n_{0}$, and depends, in particular, on whether
shortcut lengths are commensurable with the 1D ring to which the shortcuts
are added. Indeed, the obvious limiting case with $n_{0}=N/2$ is found to
maximally enhance the network \emph{traversal} time, but is associated with
the largest average access time, over all such networks with fixed average
shortcut degree $\bar{k}$.

The rest of the paper is laid out as follows. In the next section we
introduce the master equation governing the evolution of the probabilities $%
P_{m}\left( t\right) $ for the walker to be at a given site on such a
network at time $t$. For each network, the master equation can in principle
be solved numerically in either the time or frequency (Laplace) domain, and
the results averaged over an appropriate ensemble of networks to obtain
observable quantities of interest. This becomes increasingly difficult for
large $N$ due to the size of the random matrices that must be manipulated in
the process. Equivalent results can be obtained by performing Monte Carlo
simulations that follow the trajectory of individual walkers, using the
assigned transition probabilities to determine a specific random sequence of
steps and transition times. Simulation results must then be doubly averaged,
first over a sufficient number of walks, and then over an appropriate
ensemble of networks. A useful \emph{fixed jump time} simulation technique
for computing mean traversal and access times for the underlying \emph{%
continuous} time random walk is described in Sec. II for this purpose. The
averaging process underscores the point that one is interested in
statistical properties of the ensemble of networks associated with a
particular set of connection probabilities $q_{n},$ which for large enough
networks should be representative of typical properties associated with any
given realization. Section III presents an analytic approach to the problem
with the construction of an approximate, but numerically accurate, effective
medium theory (EMT) for describing the evolution of the ensemble averaged
probabilities $p_{m}\left( t\right) =\langle P_{m}\left( t\right) \rangle ,$
from which the mean traversal times can be easily and quickly computed, even
for very large networks. In Sec. IV we present and discuss the main
numerical (Monte Carlo) results of the paper, comparing them in each case
with the predictions of our analytic effective medium theory; we find them
to be generally in excellent agreement with each other. The last section of
the paper contains a summary.

\section{Network Models, Master Equations, and Simulation Techniques}

As is clear from the introduction, we are interested in random walks in
continuous time as described by the master equation
\begin{eqnarray}
\frac{dP_{m}}{dt} &=&\sum_{n=\pm 1}F\left( P_{m+n}-P_{m}\right) +\sum_
{n\neq \pm 1}f_{m,n}\left( P_{m+n}-P_{m}\right)  \notag \\
&=&-\sum_{n}A_{m,n}P_{n}  \label{ME}
\end{eqnarray}%
where on the ring it is understood that $P_{m}=P_{m+N},$ and the second sum
in the first line includes all values $n\in \left\{ \pm 2,\pm 3, \cdots ,\pm
\left( N/2-1\right) ,N/2\right\} ,$ but excludes nearest-neighbors ($N $ is
assumed even). The transition rate $f_{m,n}=f_{m+n,m}$ between node $m $ and
node $m+n$ is a random variable equal to $f$ with probability $q_{n}$ and to
zero with probability $1-q_{n}.$ In the second line we have expressed the
linear relation on the right-hand side in terms of the transition matrix $%
\mathbf{A}$. Indeed, the solution to the master equation can be numerically
obtained through an evaluation of the exponential of this matrix.
Specifically, the probability to find the walker at node $m$ at time $t,$
given that it started at node $n$ at $t=0$ is the Green's function or
propagator \cite{par05}%
\begin{equation*}
G_{n,m}\left( t\right) =\left[ e^{-\mathbf{A}t}\right] _{n,m}.
\end{equation*}%
The functions $G_{n,m}\left( t\right) $, or their Laplace transforms
\begin{equation}
\tilde{G}_{m,n}\left( \varepsilon \right) =\int_{0}^{\infty }G_{m,n} \left(
t\right) e^{-\varepsilon t}dt=\left[ \left( \varepsilon +\mathbf{A} \right)
^{-1}\right] _{m,n}  \label{gnm}
\end{equation}%
are readily computed for moderately sized systems $N\lesssim 10^{3}$ using
standard numerical techniques, but computation times and storage
requirements can grow significantly with system size $N$.

Information about traversal and access times is straightforwardly obtainable
from the Green's functions or their Laplace transforms, Eq.~(\ref{gnm}). For
example, the mean first passage time $\tau _{m,n}$ for a walker to arrive at
node $m$ given that it started at $n$ is computable through the expression %
\cite{par05}%
\begin{equation}
\tau _{m,n}=-\lim_{\varepsilon \rightarrow 0}\frac{d}{d\varepsilon } \left[
\frac{\tilde{G}_{m,n}\left( \varepsilon \right) }{\tilde{G}_{m,m}\left(
\varepsilon \right) }\right] .  \label{tauFPT}
\end{equation}%
In this way, general traversal and access times are readily computed from
the resolvent of the matrix $\mathbf{A}$ for small values of the Laplace
variable $\varepsilon $; a full solution of the problem requires only the
operation of matrix inversion.

\begin{figure}[tbp]
\epsfysize=3.5truein\epsffile{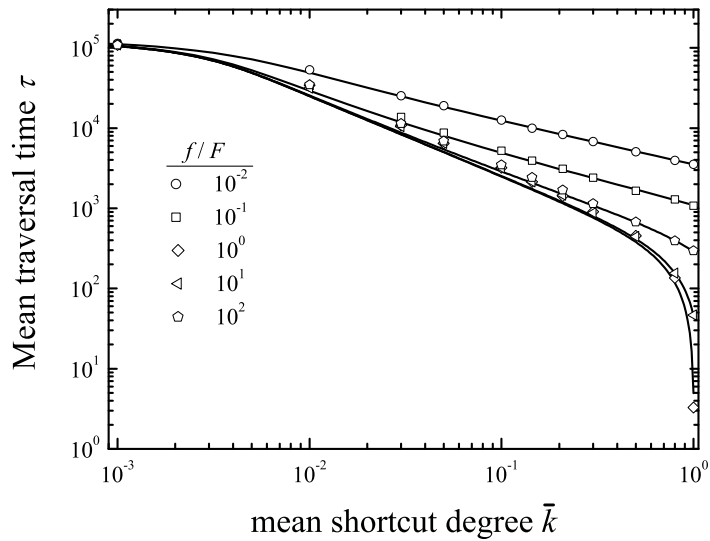}
\caption{Mean traversal time as a function of the mean shortcut degree $\bar{%
k}$ on a TEN of size $N=1000$, with links of maximal length $n_{0} = N/2$,
for different values of $f/F$, as indicated.}
\label{FIG. 1}
\end{figure}

\begin{figure}[tbp]
\centering {\epsfysize=2.3in \epsffile{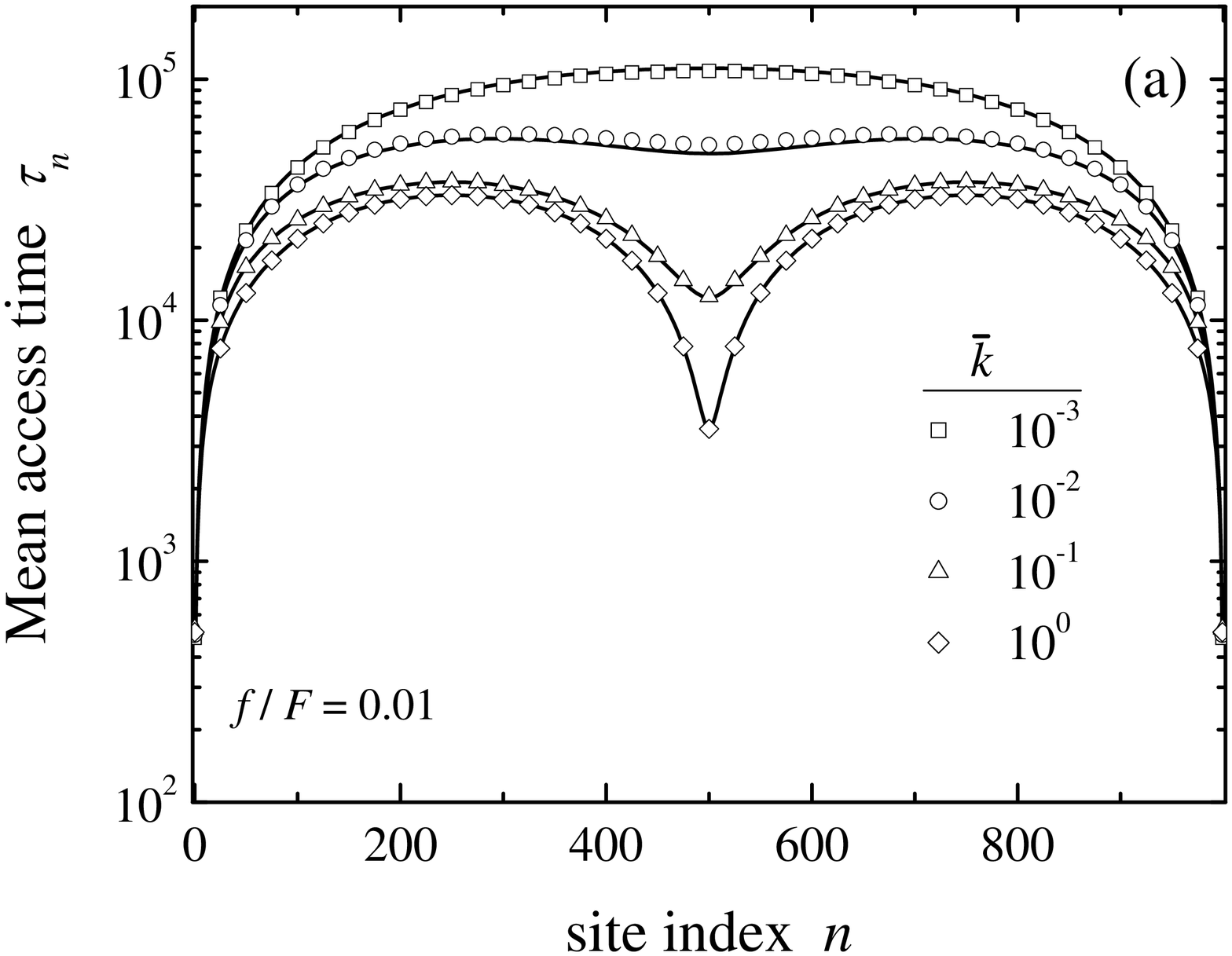}} {\epsfysize=2.3in %
\epsffile{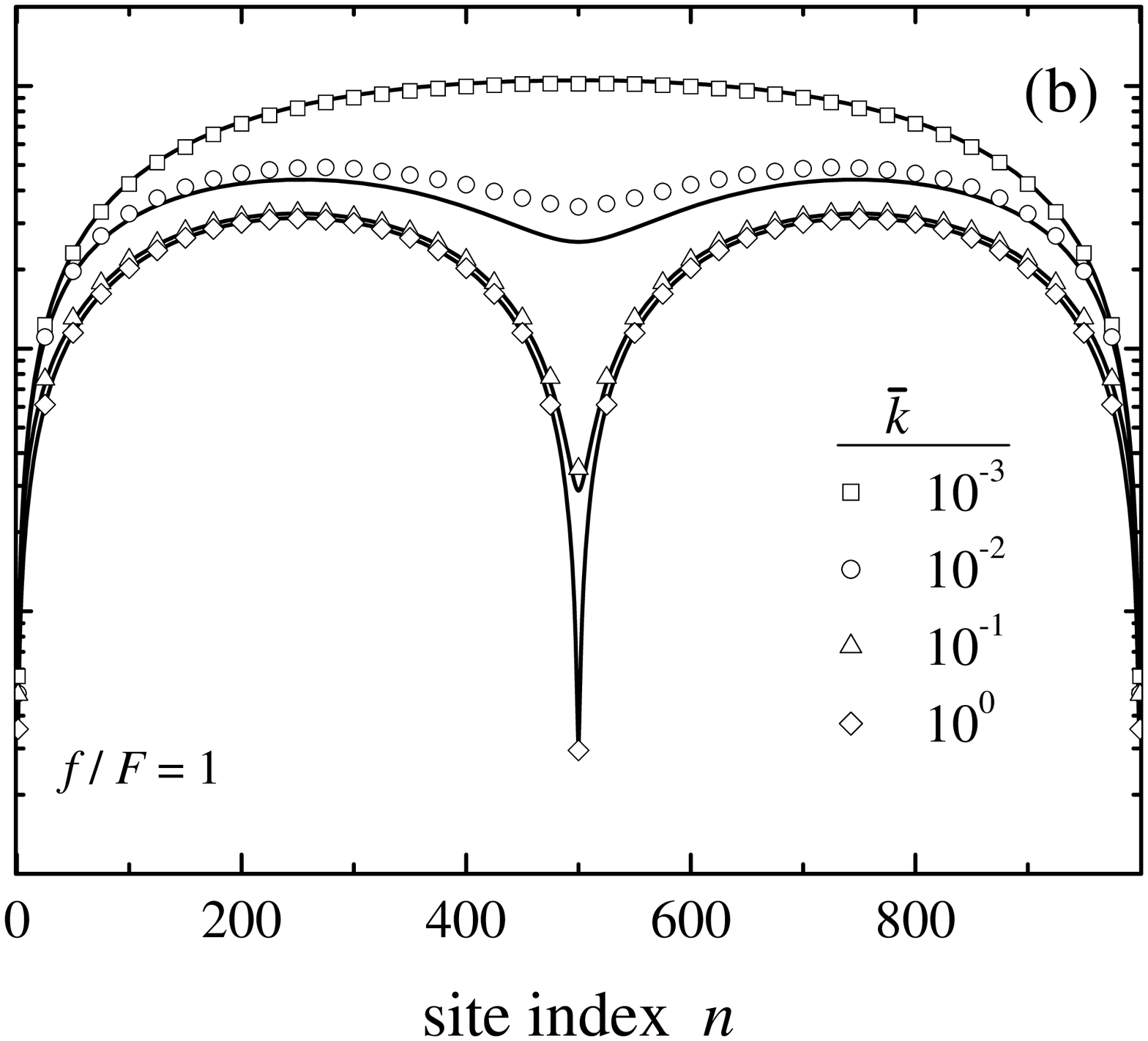}} {\epsfysize=2.3in \epsffile{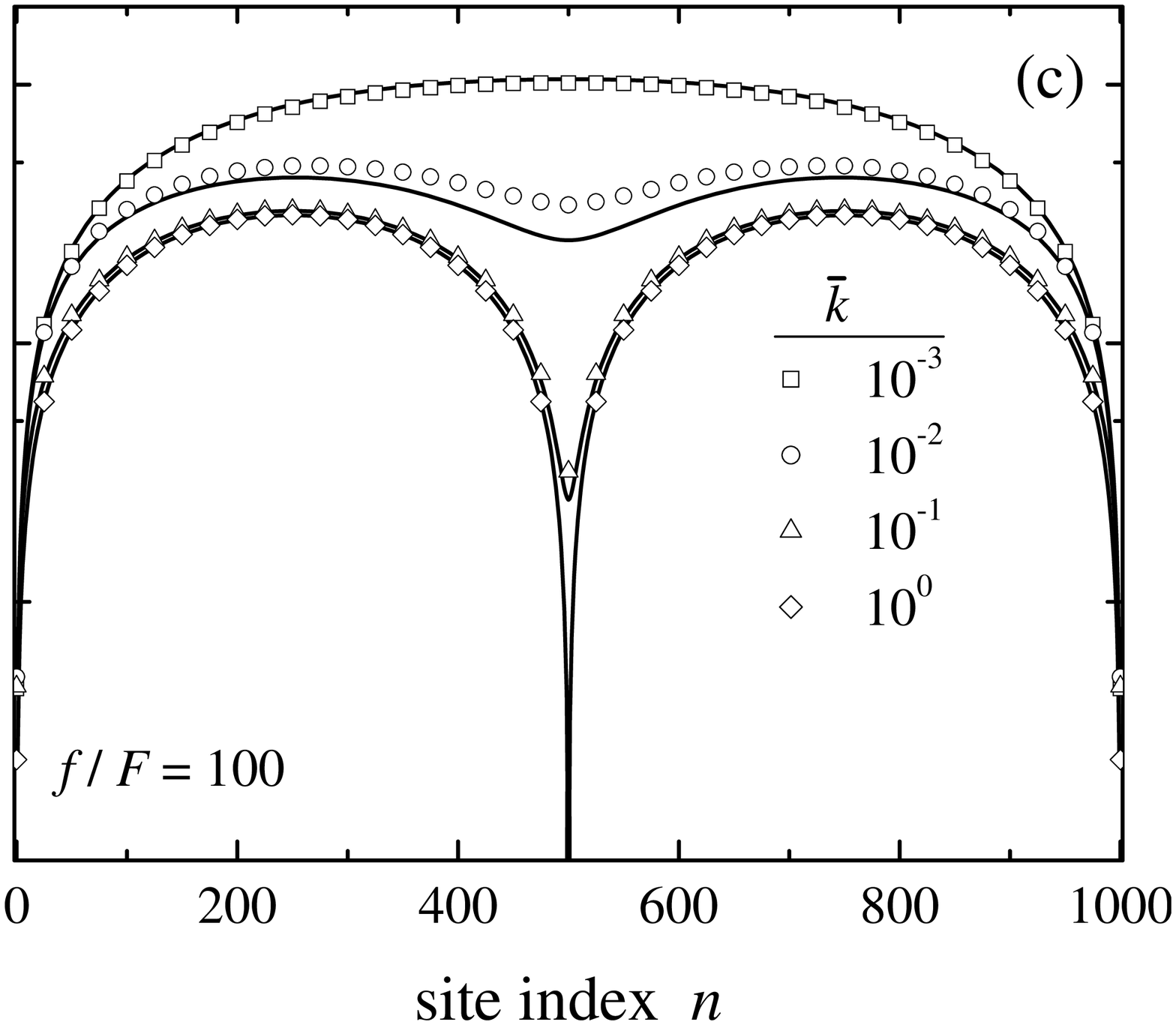}}
\caption{Access times for TENs of size $N=1000$, with $n_{0} = N/2$, as a
function of the site position $n$ (where the random walk's initial position
is $n=0$). The plots correspond to different shortcut densities in the range
$0.01\leq \bar
{k}\leq 1$, as indicated, and different jump rate ratios:
(a) $f/F=10^ {-2}$; (b) $f/F=1$; (c) $f/F=10^{2}$. Vertical scales are the
same in (a)-(c).}
\label{FIG. 2}
\end{figure}

\begin{figure}[tbp]
\center \epsfysize=3.5truein\epsffile{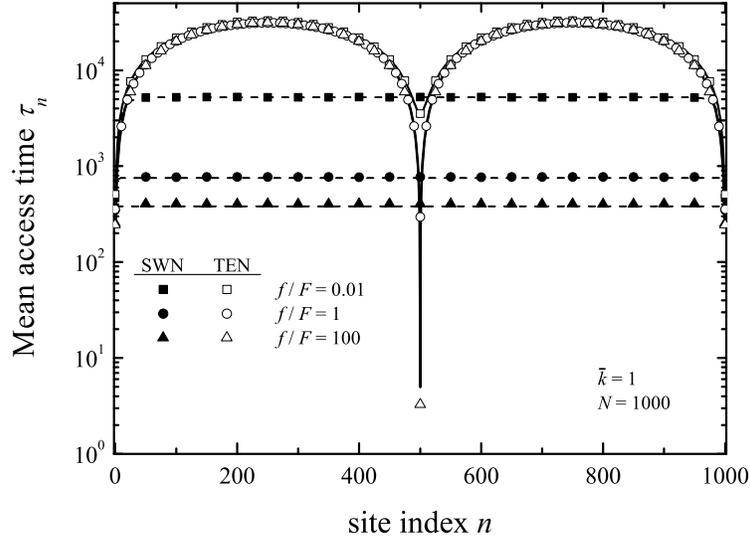}
\caption{Comparison between access times for the SWN and the TEN (with $%
n_{0} = N/2$) for $\bar{k}=1$, $N=1000$, and different jump rate ratios, as
indicated.}
\label{FIG. 3}
\end{figure}

\begin{figure}[tbp]
\center \epsfysize=3.5truein\epsffile{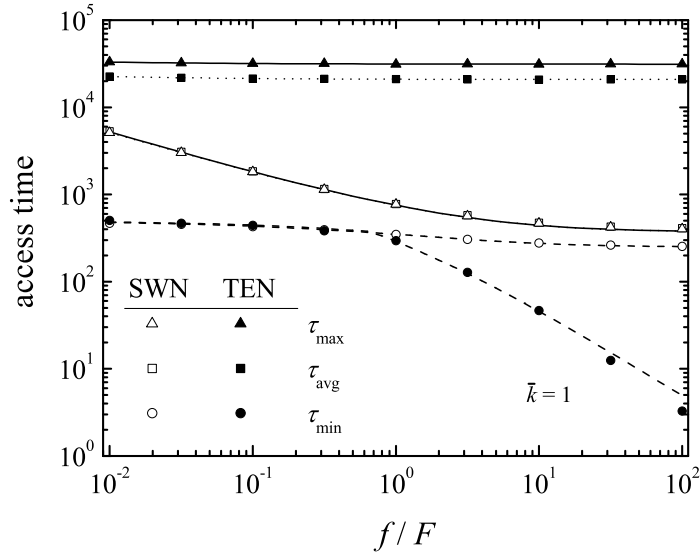}
\caption{Comparison between the minimum, average and maximum mean access
times for the SWN and the TEN, respectively. As in the previous Figure, we
consider $\bar{k}=1$, $N=1000$, and $n_{0} = N/2$.}
\label{FIG. 4}
\end{figure}

On the other hand, the full solution to the master equation contains more
information than is really necessary for computing mean access times. It is
also possible to simply perform a Monte Carlo simulation \cite{can07} that
follows individual trajectories along an ensemble of random walks, with
steps and dwell times governed by the transition probabilities assigned to a
given network. Jump destinations of the random walker are chosen at each
time step from the local transition probabilities at the site occupied by
the walker, by means of pseudo- random variables (see e.g. \cite{bin02}).
Specifically, a walker located at a given site on the network with shortcut
degree $k$ will make a transition to one of its two neighbors on the ordered
ring with probability $p_{lat}=FT$, will make a transition to one of the $k$
sites to which it is connected by a shortcut with probability $p_{sh}=fT$,
and will stay at its present position with probability $p_{stay}=1-(2F+kf)T$%
. After performing $n_{MC}$ Monte Carlo steps, the corresponding continuous
time elapsed is $\tau =n_{MC}T$. As we show in the appendix, as long as we
are interested only in computing mean first passage times (and not the
probabilities themselves, or other observables), the choice of the fixed
jump time $T$ is arbitrary, provided all transition probabilities remain
positive definite (i.e., provided $T\leq \left( 2F+k_{\max }f\right) ^{-1}$.
In the appendix we prove, specifically, that this method (when averaged over
a large enough ensemble of random walks) reproduces the mean access times of
the master equation. The mean access time $\tau _{m}$ for a given network
configuration, is then computed as the average of this quantity over random
walks on the same network, and then over a sufficiently large ensemble of
networks characterized by the same set of network parameters. Except where
explicitly noted, simulation results shown in the present paper were
obtained for networks of size $N=1000$, typically averaged over 100
different network configurations and 1000 different random walk trajectories
per configuration. Specific simulation results appear
in Figs. \ref{FIG. 1}-\ref{FIG. 8}, together
with predictions of the effective medium theory derived in the next section.

\section{Effective Medium Theory for Complex Networks}

In this section we derive analytical expressions that define an effective
medium theory (EMT) for the configuration averaged probabilities, or Green's
functions
\begin{equation}
\langle \tilde{G}_{n,m}\rangle =\langle \tilde{G}_{n+\ell ,m+\ell }\rangle =%
\tilde{g}_{n-m}\left( \varepsilon \right)  \label{EMAG}
\end{equation}%
that describe transport on the very general class of structures described in
the introduction. We also specialize it to the analysis of partially-ordered
networks having a sharp distribution of link lengths, in which form it
becomes very simple to implement. As suggested by Eq.~(\ref{EMAG}) above,
the key simplification that arises in EMT is that \emph{average} transport
properties of the ensemble are translationally invariant. In the present
context, we assume that the functions $\tilde{g}_{m}\left( \varepsilon
\right) $ obey the equations%
\begin{equation}
\varepsilon \tilde{g}_{m}-\delta _{m,0}=\sum_{n=\pm 1}F\left( \tilde{g}%
_{m+n}-\tilde{g}_{m}\right) +\sum_{n\neq \pm 1}\tilde{w}_{n}\left(
\varepsilon \right) \left( \tilde{g}_{m+n}-\tilde{g}_{m}\right) .
\label{EMA ME}
\end{equation}%
Here, the $\tilde{w}_{n}\left( \varepsilon \right) $ are frequency-
dependent rates, equivalently, memory functions in the Laplace domain,
connecting pairs of sites on the network separated by a distance $\left|
n\right| $. Notice that, without loss of generality, we consider that the
random walker's initial position is at $m=0$. In the EMT equations,
transport \emph{around} the ring edge is characterized by the same rate $F$
that obtains throughout the ensemble, but the $\tilde{w}_{n}\left(
\varepsilon \right) $ must be determined from self-consistent
considerations. The \emph{simplest} (non-self-consistent) approximation,
i.e, $\tilde{w}_{n}\left( \varepsilon \right) =\langle f_{m,n}\rangle
=q_{n}f $, generally gives a very poor approximation to the dynamics, and is
equivalent to an ``annealed model'' in which possible shortcut connections
are determined anew at each step of the random walk \cite{par05}, a
situation very different from the actual ``quenched disorder'' model
considered here, which has percolative possibilities lacking in an annealed
model.

Note that once the effective medium parameters $\left\{ \tilde{w}_{n} \left(
\varepsilon \right) \right\} $ are determined, the rest of the problem is
straightforward, since the translationally invariant set of equations (\ref%
{EMA ME}) is easily solved by introducing discrete Fourier transforms $%
\tilde{g}^{k}\left( \varepsilon \right) =\sum_{m}\tilde{g}_{m}\left(
\varepsilon \right) e^{-ikm},$ which diagonalize the EMT transition matrix,
and lead to the solution%
\begin{equation}
\tilde{g}_{m}\left( \varepsilon \right) =\frac{1}{N}\sum_{k}\frac{e^ {ikm}}{%
\varepsilon +2F\cos \left( k\right) +\omega _{N/2}\left[ 1-\cos \left(
kN/2\right) \right] +\sum_{n=2,\frac{N}{2}-1}\tilde{w}_{n}\left( \varepsilon
\right) \left[ 1-\cos \left( kn\right) \right] }  \label{ksum}
\end{equation}%
where the outer sum is over all wavevectors $k=2\pi \ell /N,$ with $\ell \in
\left\{ 1,\cdots ,N\right\} ,$ where in the denominator we have separated
out the single $n=N/2$ term, for which there is only one link, and where in
the other terms (with $n<N/2$) contributions from both links of a given
length have been combined together. From these effective medium propagators
it is also straightforward to evaluate access times. Indeed, analysis of the
propagators (\ref{ksum}) reveals that the mean access times (\ref{tauFPT})
in the effective medium can be simply calculated
\begin{equation}
\tau _{m}=-\lim_{\varepsilon \rightarrow 0}\frac{d}{d\varepsilon }\left[
\frac{\tilde{g}_{m}\left( \varepsilon \right) }{\tilde{g}_{0}\left(
\varepsilon \right) }\right] =N\tilde{\gamma}_{m}\left( 0\right)
\label{taugamma}
\end{equation}%
directly from the zero frequency value of the propagator difference
\begin{equation}
\tilde{\gamma}_{m}\left( \varepsilon \right) \equiv \tilde{g}_{m}\left(
\varepsilon \right) -\tilde{g}_{0}\left( \varepsilon \right) .  \label{gamma}
\end{equation}

To determine the function $\tilde{w}_{n}\left( \varepsilon \right) $
self-consistently, we imagine that we have already performed the average
over all links \emph{except} the one connecting the origin to the site $n$
nodes away, so that the resulting equations of motion are not the completely
averaged EMT equations (\ref{EMA ME}), but are, instead, of the form%
\begin{eqnarray}
\varepsilon \tilde{p}_{m}-\delta _{m,0} &=&\sum_{n=\pm 1}F\left ( \tilde{p}%
_{m+n}-\tilde{p}_{m}\right) +\sum_{n\neq \pm 1}\tilde{w}_{n}\left(
\varepsilon \right) \left( \tilde{p}_{m+n}-\tilde{p}_{m}\right)  \notag \\
&&+\left( \delta _{m,0}-\delta _{m,n}\right) \left( f_{0,n}-\tilde{w}%
_{n}\left( \varepsilon \right) \right) \left( \tilde{p}_{n}-\tilde{p}%
_{0}\right)
\end{eqnarray}%
where $f_{0,n}$ is an actual rate ($f$ or $0$) drawn from the distribution
governing it. For a given value of $f_{0,n}$ the solution to this defect
problem can be formally written in terms of the propagators $\tilde{g} _{n}$
for the effective medium. For example, the Laplace transformed probability
to find the walker at the origin can be expressed in terms of the EMT
quantities $\tilde{\gamma}_{n}$ in (\ref{gamma}) through the relation%
\begin{equation}
\tilde{p}_{0}=\tilde{g}_{0}-\frac{\left( f_{0,n}-\tilde{w}_{n}\right) \tilde{%
\gamma}_{n}^{2}}{1+2\left( f_{0,n}-\tilde{w}_{n}\right) \tilde{\gamma} _{n}}.
\end{equation}%
Averaging this over the binary distribution associated with this one
remaining link must now generate the \emph{completely} averaged effective
medium, i.e., we self-consistently demand that $\langle \tilde{p}_{0}\rangle
=\tilde{g}_{0}.$ Hence the average of the second term on the right hand side
of this last equation must vanish. Repeating for each node $n $ gives the
following set of self-consistent conditions%
\begin{equation}
\frac{q_{n}\left( f-\tilde{w}_{n}\right) }{1+2\left( f-\tilde{w}_{n} \right)
\tilde{\gamma}_{n}}=\frac{\left( 1-q_{n}\right) \tilde{w}_{n}}{1-2 \tilde{w}%
_{n}\tilde{\gamma}_{n}}\qquad \qquad n=2,\cdots ,N/2,  \label{SCone}
\end{equation}%
which are equivalent to the relations $2\gamma _{n}\tilde{w}_{n}^{2}- \tilde{%
w}_{n}\left( 1+2f\gamma _{n}\right) +q_{n}f=0.$ By (e.g., numerically)
solving this set of simultaneous equations at each value of $\varepsilon ,$
the set of functions $\left\{ \tilde{w}_{n}\left( \varepsilon \right)
\right\} $ can be determined, and the results used to evaluate the effective
medium propagators, and thus the traversal and access times. Note that, in
general, an analytic solution is not available since the functions $\tilde{%
\gamma}_{n}\left( \varepsilon \right) $ themselves depend in a complicated
way on the complete set of effective medium parameters $\left\{ \tilde
{w}%
_{n}\left( \varepsilon \right) \right\} ,$ through (\ref{ksum}) and (\ref%
{gamma}).

Note also that in the SWN hybrids studied in Ref.~\cite{par05}, where $%
q_{n}=q$ for all shortcuts, and where a single effective medium parameter $%
\tilde{w}_{n}=\tilde{\omega}$ was assumed, the value of $\tilde {\omega}$
was identified as the root of the equation
\begin{equation}
\sum_{n=2}^{N/2}\frac{q\left( f-\tilde{\omega}\right) }{1+2\left( f- \tilde{%
\omega}\right) \tilde{\gamma}_{n}}=\sum_{n=2}^{N/2}\frac{\left( 1-q \right)
\tilde{\omega}}{1-2\tilde{\omega}\tilde{\gamma}_{n}}\qquad \qquad n=2,
\cdots ,N/2
\end{equation}%
obtained by summing (or averaging) each of the equations (\ref{SCone}). In
the present context, we anticipate generally having to simultaneously solve
the set of $\left( N-2\right) /2$ equations (\ref{SCone}) for the
corresponding number of effective medium parameters $\tilde{w}_{n}.$

Of course, in situations in which $q_{n}=0,$ the vanishing of the left hand
side of (\ref{SCone}) leads to a vanishing of the associated effective
medium rate $\tilde{w}_{n}\left( \varepsilon \right) $ as well, and hence a
reduction in the number of coupled equations that must be solved. Thus, in
the simplest such network, for which $q_{m}=q\delta _{m,n_{0}},$ and in
which only shortcuts of a single fixed length $n_{0}$ are added to the ring,
the set reduces to one equation for determining the single non-zero
effective medium parameter $\tilde{w}_{n_{0}}\left( \varepsilon \right) ,$
which is easily determined numerically, and from which the corresponding
access times can then be calculated using (\ref{taugamma}). In the next
section we present the results of Monte Carlo simulations on such networks
along with the corresponding predictions of the effective medium theory
derived above.

\section{Numerical Results and Discussion}

In Fig.~\ref{FIG. 1} we show the mean traversal time $\tau =\tau _{N/2}$
(normalized to the mean hopping time $\tau _{F}=F^{-1}$ for hops around the
edge of the ring) as a function of the mean shortcut degree $\bar{k}$ on a
TEN of $N=1000$ sites with links of maximal length $n_{0}=N/2,$ for
different values of $f/F$. Open symbols are the results of simulation, the
solid curves running near or through them are the predictions of EMT. As
expected, the mean traversal time decreases monotonically as the mean degree
of the network increases, since a larger density of shortcuts naturally
shortens the time needed to cross the network. The decrease in traversal
time is modest for small values of the ratio $f/F$, but becomes quite large
for $f\geq F$ and for shortcut densities close to the maximum value, which
for this network occurs when the mean shortcut vertex degree $\bar{k}=\bar
{%
k}_{max}=1$. The decrease seen in Fig.~\ref{FIG. 1} is quite similar to that
previously encountered in the SWN case \cite{par05}, where, for low shortcut
densities and $f\geq F$, the data were found to collapse onto a universal
curve, while, for higher shortcut densities near $\bar{k}\simeq 1$,
traversal times scaled as $\tau \sim (f/F)^{-1}$. In this regime of the SWN,
as in the present case, by increasing the ratio $f/F$ the mean traversal
time can be made arbitrarily small \cite{par05}. Not surprisingly, the
behavior seen in the SWN and in the TENs of Fig.~\ref{FIG. 1} arise for a
similar reason. In the limit $\bar{k}\ll 1$, the shortcut density is very
low and the traversal time tends to the diffusive limit, where $2F\tau
_{diff}=(N/2)^{2}$. Even when $f/F\gg 1$, if $\bar{k}\ll 1$, the shortcut
connections are very sparse, and there will not be a substantial reduction
in the traversal time, as the walker still spends most of its time on the
slower steps it must take \emph{around} the ring. In this regime, therefore,
the traversal time is largely insensitive to the value of $f/F,$ as long as
it is large enough for the shortcuts not to be rate limiting. However, when
the number of shortcuts approaches a threshold near $\bar{k}\simeq 1$,
percolation of the underlying shortcut network leads to a change in the
transport mechanism; in this regime particles need few if any slow steps to
get across the system, and can move mainly along the fast shortcut
connections.

In order to further compare characteristic features of transport on
different network structures, we also consider the distribution and spatial
variation of access time $\tau _{n}$ for going from an arbitrary initial
site to one located $n$ nodes away along the 1D backbone. Figure~\ref{FIG. 2}
shows access times $\tau _{n}$ as a function of node position $n$ for the $%
n_{0}=N/2$ TEN at different shortcut densities in the range $1\geq \bar{k}%
\geq 0.01,$ as indicated, and for different jump rate ratios: (a) $%
f/F=10^{-2}$, (b) $f/F=1$, and (c) $f/F=10^{2}$. Again, open symbols
indicate simulation results, and solid lines the predictions of the EMT,
which is generally quite good, except in a few cases where it tends to
underestimate the access time. For small mean shortcut degrees, $\bar
{k}%
\ll 1$, access times grow monotonically with distance, as transport is
dominated by diffusion along the regular lattice bonds. For denser networks
on the other hand, one observes a local minimum taking place at $n=n_{0}=N/2$%
, reflecting the traversal time enhancement that might be expected of the
TEN structure.

Figures~\ref{FIG. 2}(a)-(c) show the effect of increasing the speed of
shortcut connections. As expected, in the sparse network limit ($\bar
{k}%
\ll 1 $) access times are essentially universal and are insensitive to
shortcut speeds, while for denser networks access is enhanced dramatically
around $n=n_{0}=N/2$ for large values of $f/F$. Indeed, at $f/F=10^{2}$ (see
Fig.~ \ref{FIG. 2}(c)), an improvement of almost 3 orders of magnitude in $
\tau _{N/2}$ is achieved by increasing the shortcut density from $\bar{k}
=0.1$ to $\bar{k}=1$. The improvement is dramatic, but \emph{not} very
democratic, and is largely confined to a region in the immediate vicinity of
$n_ {0}$.

\begin{figure}[tbp]
\center {\epsfysize=1.9in \epsffile{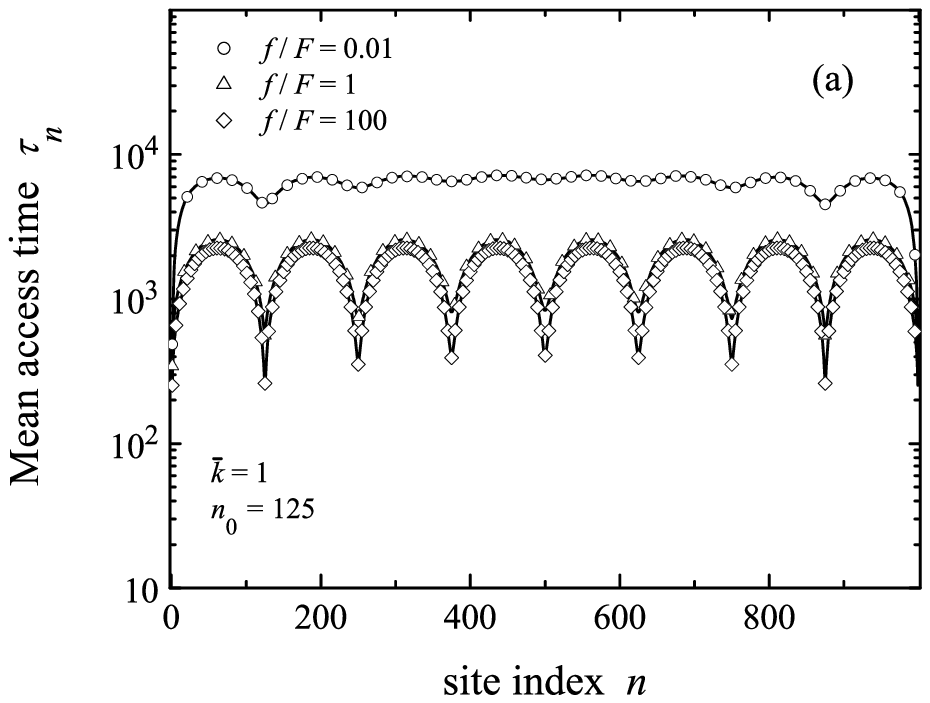}} {\epsfysize=1.9in
\epsffile{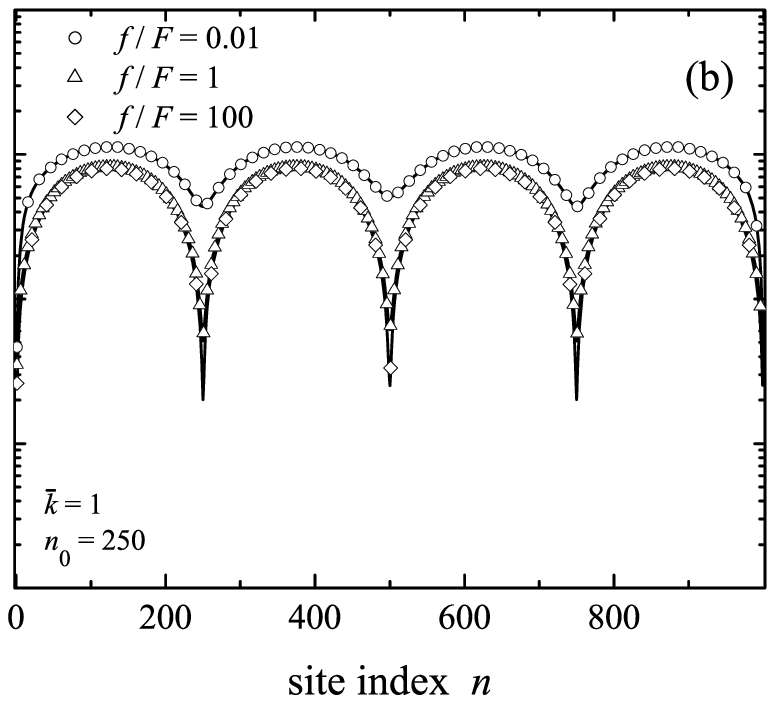}} {\epsfysize=1.9in \epsffile{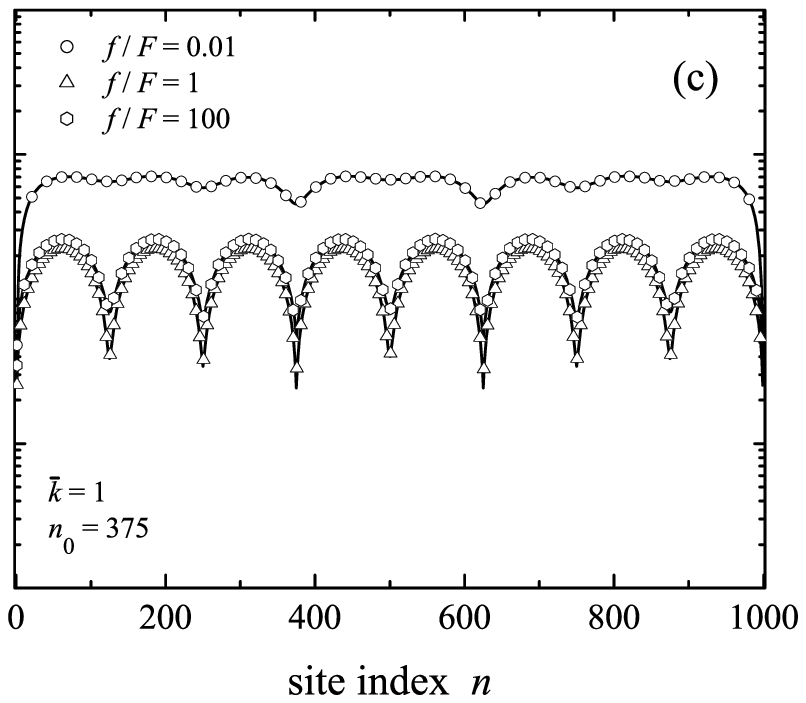}}
\caption{Access times for TEN structures of size $N=1000$, mean degree $\bar{%
k}=1$, and shortcuts of different length: (a) $n_{0}=125$, (b) $n_{0} =250 $%
, and (c) $n_{0}=375$. Vertical scales are the same in (a)-(c).}
\label{FIG. 5}
\end{figure}

To highlight this point, we show in Fig.~\ref{FIG. 3} a comparison of access
times for TEN and SWN structures, for $\bar{k}=1$ and for different jump
rate ratios $f/F$, as indicated \cite{footnote1}. In this figure we see a
sharp contrast between access times in TEN structures (which clearly reflect
their specific design goal of improving access to sites directly across the
network) and SWN access time distributions, which are remarkably flat, and
which, for the same number of links, do a better job of providing faster
access times to \emph{most} network nodes.

Further insights on the different transport properties of TEN and SWN
structures can be gained by considering simple statistical properties of the
access time distribution, such as the minimum, average, and maximum access
times, defined by $\tau _{min}=\mathrm{\min }\{\tau _{n}\}$, $%
\tau_{avg}=(N-1)^{-1}\sum_{n}\tau _{n}$ and $\tau _{max}=\mathrm{\max }
\{\tau_{n}\}$, respectively.

From our previous discussion regarding the results in Fig. \ref{FIG. 2}, it
should be noted that, for TENs in the parameter region $\bar{k}\simeq 1$ and
$f/F\gg 1$, the mean \emph{traversal} time is actually the minimum access
time for the network, i.e. $\tau _{min}\sim \tau _{N/2}$. For SWNs, on the
other hand, the mean and maximum access times are essentially the same, and
well approximated by the traversal time: $\bar{\tau}\sim \tau _{max}\sim
\tau _{N/2}$. A comparison of $\tau _{min}$, $\tau _{avg}$ and $\tau _ {max}$
for TENs and SWNs as a function of the rate ratio $f/F$ is shown in Figure %
\ref{FIG. 4}. As in Fig. \ref{FIG. 2}, we consider dense networks with $\bar{%
k}=1$. For both types of network, when $f/F\leq 1$ the minimum access time $%
\tau _{min}$ corresponds to the time to arrive at one of its nearest
neighbors. However, in the TEN case a crossover takes place around $%
f/F\approx 1$ to a regime in which $\tau _{min}$ corresponds to the
traversal time for going all the way across the network. Indeed, for $f/F\gg
1$ it is observed that $\tau _{min}\ll \tau _{avg},\tau _{max}$, reflecting
the large traversal enhancement effect characteristic of TENs in this
parameter region. The differences among $\tau _{min}$, $\tau _{avg},$ and $%
\tau _{max}$ are much less pronounced in the case of SWNs, which is again a
manifestation of the nearly flat distribution of this type of network.

In order to investigate further the properties of TEN structures, in
particular, we now consider the case in which shortcut connections have
lengths different from the ring's diameter $d$. If we consider shortcuts
connecting sites separated by a distance $n_{0}$ (measured along the
backbone ring in units of lattice bond length) as straight rods or wires of
length $L$ , then%
\begin{equation}
L=d\;\mathrm{\sin }\left( \pi n_{0}/N\right) .  \label{shlength}
\end{equation}

\begin{figure}[tbp]
\center{\epsfysize=2.3in \epsffile{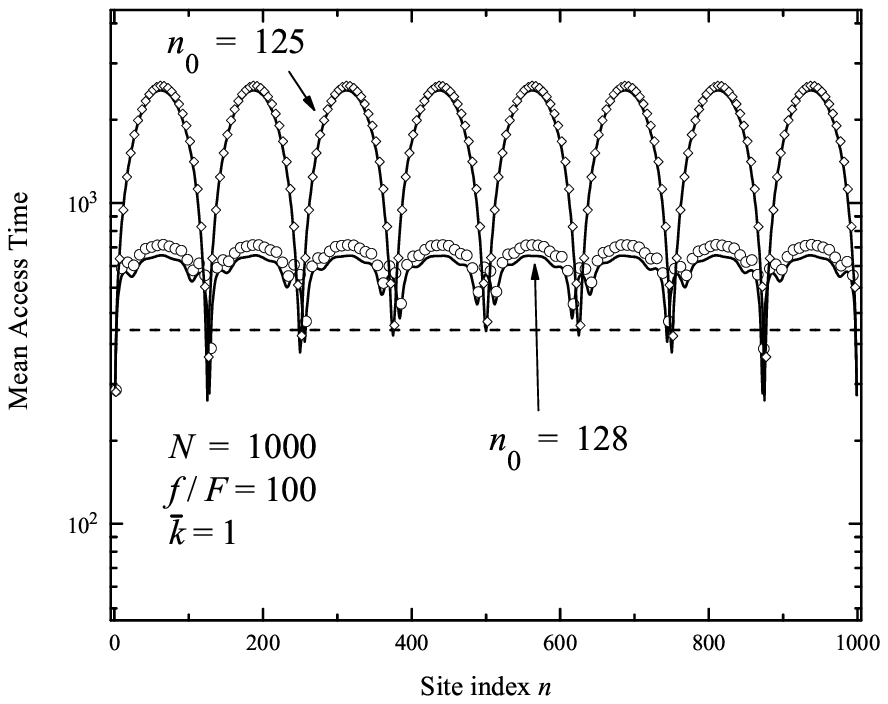}}
{\epsfysize=2.3in\epsffile{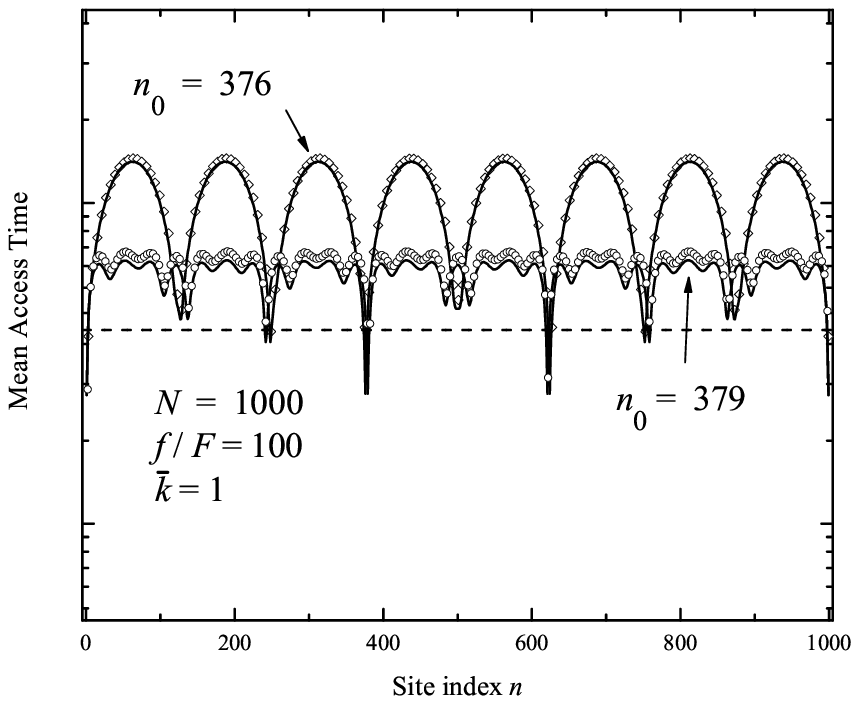}}
\caption{Access times for TEN structures of size $N=1000$ and rate ratio $%
f/F=100$. Results for connections of length $n_{0}=125$ and $128$ appear in the left panel, and of length $n_{0}=376$ and $379$ in the right panel. Symbols are the results of numerical simulations, solid lines the predictions of effective medium theory. Horizontal dashed line is the mean access time for SWNs at the same shortcut density.}
\label{FIG. 6}
\end{figure}

\begin{figure}[tbp]
\center \epsfysize=3.5truein\epsffile{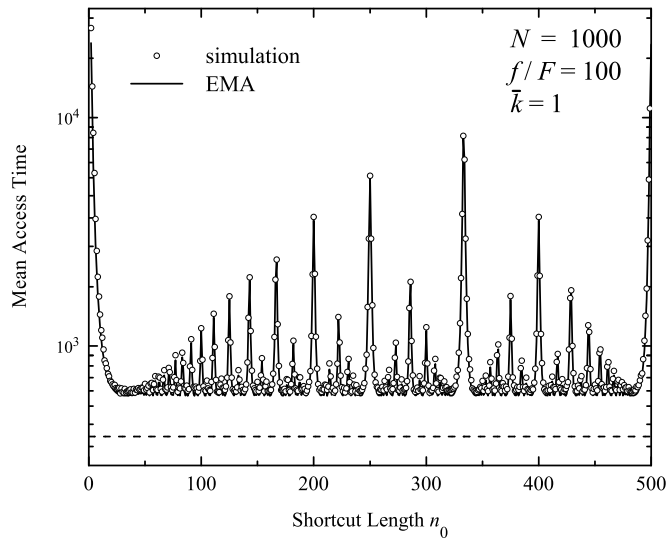}
\caption{Average network access times for TEN structures of size $N=1000$ and rate ratio $%
f/F=100$, as a function of the length $n_{0}$ of the shortcuts, at fixed shortcut density. Horizontal dashed line is the mean access time for SWNs at the same shortcut density.}
\label{FIG. 7}
\end{figure}

Figure \ref{FIG. 5} shows access times for TEN structures of $N=1000$ sites
with mean degree $\bar{k}=1$, and shortcuts of different length: (a) $%
n_{0}=125$, (b) $n_{0}=250$, and (c) $n_{0}=375$. Here the mean vertex
degree is now one half the maximum value it can attain, since there are now
two possible links of length $n_{0}$ connected to any given site. For these
structures, an interesting oscillatory pattern arises for $f/F\gg 1$, with a
characteristic repeat length related to the shortcut length $n_{0}$ and to
the commensurability of $n_{0}$ with the total number of sites in the ring.
Notice the similarity between the patterns for $n_{0}=125$ and $n_{0} =375$,
which presumably arises due to the fact that when one moves around the ring
with steps of length $n_{0}=125,$ the path closes after one circuit, and
revisits the same set of 8 sites, separated by a distance $n_{0}$, while
with steps of length $n_{0}=375,$ the path closes after two circuits, over
which it visits the \emph{same} set of sites (modulo $1000$)\ as visited in
one circuit with steps of length $n_{0} = 125.$ Like the system in which $n_{0} = N/2$, however, there are clearly significant regions of the network for which the speed-up in access time is not very significant. Are such gaps inevitable? Perhaps not. In these examples the shortcut length is commensurate with the circumference of the ring. One might imagine, based upon the description above for $n_{0}=375,$ that the periodic structure might be ``filled in'', by choosing a step length that is highly
incommensurate with the circumference of the ring. Indeed, an investigation of this effect confirms that a slight alteration in the shortcut length can lead to dramatic changes in the spatial distribution of mean access times. We display this effect in Fig. \ref{FIG. 6}, where we show the mean access time distribution for four systems, consisting of two pairs with shortcut lengths that differ by a relatively small number of sites. The dramatic change in the shape, and the overall mean access time, suggests that the latter quantity is a strongly fluctuating function of the shortcut length $n_{0}$ for this class of networks. In Figure \ref{FIG. 7} we present a plot showing this strong variation of the mean access time of the entire network as a function of the shortcut length $n_{0}$, for a fixed average number of shortcuts. In Figs. \ref{FIG. 6} and \ref{FIG. 7} the solid lines are the result of the effective medium theory, the symbols the result of numerical simulation. Effective medium theory does an excellent job capturing the highly fluctuating access time spectrum. The results show that in addition to speeding up access times to the specific sites connected by shortcuts of a given chosen length, a significant reduction in the overall access time is also possible, provided that the shortcut length is not commensurate with existing structures in the network to which the shortcuts are added.

\begin{figure}[tbp]
\center \epsfysize=3.5truein\epsffile{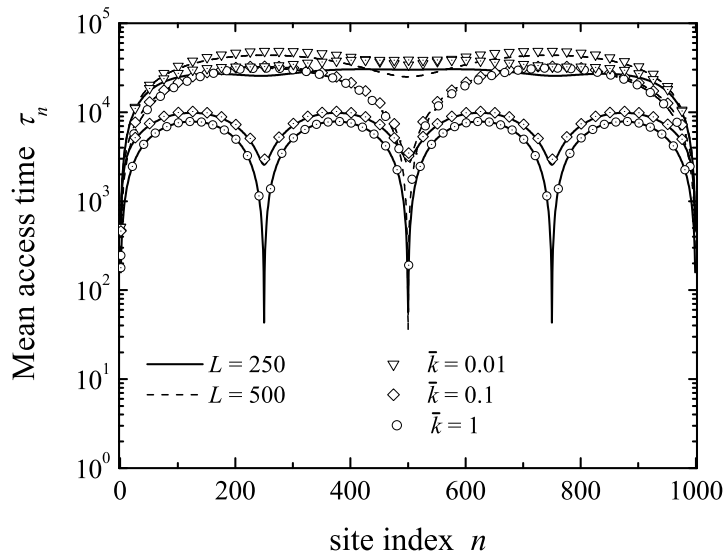}
\caption{Average network access times for TEN structures of size $N=1000$ and rate ratio $%
f/F=100$, and different shortcut densities are compared for systems with the same total length of all shortcuts.}
\label{FIG. 8}
\end{figure}

Finally, as discussed in the introduction, one of the main interests of studying transport properties on different network topologies (with the addition of shortcuts to an underlying ordered structure) is aimed at optimizing access times and
signal transmission across the network, where added shortcuts might
represent fast, high quality connections ($f/F\gg 1$). In this context, let
us consider a comparison at fixed ``cost'' (as represented by the total
length of shortcut wires, i.e. $L_{total}=L\times (N\bar{k}/2)$) and compare
access times obtained on TEN configurations for different values of $L $. As
an example, Fig. \ref{FIG. 8} shows access times for two TEN structures of
size $N=1000$, where fast connections ($f/F=100$) of length $n_{0}=250 $ and
$n_{0}=500$ are compared for corresponding values of the shortcut densities
that keep $L_{total}$ fixed. From Eq.(\ref{shlength}) and the condition of
equal total length, it turns out that $\bar{k}(n_{0}=N/4)=\sqrt{2}\times
\bar{k}(n_{0}=N/2)$. One can see that the use of a large number of shorter
connections is generally more effective in reducing access times to most of
the network nodes, except for sites located in the immediate neighborhood of
the region spanned by the longer connections.

\section{Summary}

In this paper we have investigated the transport of walkers in a class of
networks similar to the Newman-Watts small world networks on the basis of an
effective medium theory. The primary motivation has been to study the effect
of varying the shortcut distribution in a well-defined way. The intuition
that replacing shorter links in a small world network with an equal number
of longer ones might improve access times across the network is only
\emph{partially verified}. In keeping with
straightforward intuition, it has been
found that, if the goal is to simply improve access between nodes lying at
specific distances, then randomly adding links of that length will indeed
improve the specific subset of access times between nodes located at those
distances, sometimes dramatically so. In cases where the shortcut lengths are commensurable with the underlying size of the system, this reduction in access times is not generally reflected in the access time of the network as a whole, with large regions of the network retaining surprisingly large access times. In situations where the shortcut length is highly incommensurate with the size of the system, however, the reduction in access times expected at distances associated with the shortcut length are also at least partially shared by most of the remaining sites on the network. Presumably this would reflect the more typical behavior that one would expect in adding shortcuts of a fixed length to an already disordered structure. It also suggests a strategy for probing hidden structures in apparently disordered networks.

\appendix*
\section{}
In this appendix we show how the numerical procedure described in the body of
the article is able to correctly compute mean access and traversal times,
using a time step $T$ that is chosen arbitrarily, provided the associated
transition probabilities remain positive. We hasten to point out that if we
were interested in numerically simulating (or integrating) the Master
equation (\ref{ME}), such an approach would not work: for numerical accuracy the
time step $T$ would have to be chosen much smaller than the smallest hopping
time in the system. On the other hand, when enough walks are employed the
process described above does correctly compute mean access times.

To see this, we consider a walker that has just arrived at some site $m$ on
the network. This site will have $2$ neighbors on the ring connected to it
by transition rates $F$, and $k_{m}$ shortcut sites to which it is connected
with rates $f.$ From the master equation it is well known, and
straightforward to show, that a walker will leave such a site at a random
time $\tau $ drawn from the exponential jump time probability distribution%
\begin{equation*}
\psi _{m}\left( \tau \right) =\Gamma _{m}e^{-\Gamma _{m}\tau }\qquad \qquad
\Gamma _{m}=2F+k_{m}f.
\end{equation*}%
When it does make a transition, the probability $p_{s}$ for it to go to one
of the shortcut sites to which it is connected, and the probability $p_{R}$
for it to go to one of its two neighboring sites on the ring, are given by
\begin{equation*}
p_{s}=\frac{f}{2F+k_{m}f}\qquad \qquad p_{R}=\frac{F}{2F+k_{m}f}
\end{equation*}%
In the numerical procedure described in Sec. II, these branching ratios are
preserved each time a jump occurs, as is easily verified. Thus, each random
walk \emph{path} generated by our numerical procedure has the same
statistical weight as it would if governed by the exact jump time
distribution given above. It suffices therefore to consider the ensemble of
random walks that take place \emph{on any given random walk path}, and to
show that our numerical procedure correctly predicts the correct average
duration of walks occuring on such a path. If the initial site and final
site of such a path are taken to be the ones whose access time is of
interest, then such a demonstration is equivalent to showing that the
numerical procedure correctly calculates these quanitities of interest.

We note, however, that the average duration of walks occuring on a
particular path is just the sum of the average times required for each step
along the way. The average time that the particle waits at a given site $m,$
as described above, before hopping is%
\begin{equation*}
\tau _{m}=\int_{0}^{\infty }d\tau \;\tau \psi _{m}\left( \tau \right)
=\Gamma _{m}^{-1}.
\end{equation*}%
In the numerical procedure described in Sec. II, on the other hand, the
hopping time distribution function is not a continuous exponential
distribution, but a discrete one, i.e., the particle can only leave site $m$
at integral multiples $T_{\ell }=\ell T$ of the basic time step $T.$ The
probability $p_{\ell }$ that it does so at the $\ell $th such step is
\begin{equation*}
P_{\ell }=p^{\ell -1}q
\end{equation*}%
where $p=p_{stay}=1-(2F+k_{m}f)T$ and $q=1-p=(2F+k_{m}f)T.$ The average time
that it spends at this site, with the numerical procedure we have
implemented, is then easy to compute. We find
\begin{equation*}
\tau _{m}=\sum_{\ell =1}^{\infty }\ell T\;P_{\ell }=\left( 2F+k_{m}f\right)
^{-1}=\Gamma _{m}^{-1},
\end{equation*}%
which is the same as for the (correct) continuous exponential jump time
distribution function, and that it is, as claimed, independent of the size
of the time step, provided that all transition probabilities remain
positive. It is clear that for the computation of access times, one could
implement any hopping time distribution that has the correct mean pausing
time at each site.

\section*{Acknowledgments}

We acknowledge useful discussions with Birk Reichenbach at the beginning of
this research. This work was supported in part by the NSF under grant nos.
INT-0336343, ITR DMR-0426737 and CNS-0540348 within the DDDAS program, and
by the James S. McDonnell Foundation.  P.E.P. acknowledges the hospitality of the University of New Mexico Consortium of the Americas for Interdisciplinary Science, and Department of Physics and Astronomy, for the period over which this research was performed.


\begin{thebibliography}{99}
\bibitem{footnote1} Notice that the effects of adopting different network
structures are more noticeable for large shortcut densities (close to $\bar{k%
}\simeq 1$), since otherwise the behavior tends to the universal diffusion
limit previously discussed.
\end{thebibliography}
\end{document}